\documentclass{aa}
\usepackage{morefloats}
\usepackage{epsfig}
\usepackage{natbib}
\begin{document}
\def\kms{km~s$^{-1}${}}

\title{Long slit spectroscopy of a sample of isolated spirals with and 
without an 
AGN. \thanks{Based on observations made with WHT operated on the island of
La Palma by ING in the Spanish Observatorio del Roque de Los Muchachos
of the Instituto de Astrof\'\i sica de Canarias, the European Southern
Observatory (La Silla), Calar Alto Observatory (Almer\'{\i}a, Spain)
and Las Campanas Observatories (Chile).}  }
\author{
  I.~M\'arquez \inst{1}
\and
  F.~Durret \inst{2}
\and
  J.~Masegosa \inst{1}
\and
  M.~Moles \inst{1}
\and
  J.~Varela \inst{1}
\and
  R.M.~Gonz\'alez Delgado  \inst{1}
\and
  J.~Maza \inst{3}
\and
  E.~P\'erez \inst{1}
\and
 M.~Roth \inst{4}
}
\offprints{I. M\'arquez (\sl{isabel@iaa.es}) }
\institute{
    Instituto de Astrof\'\i sica de Andaluc\'\i a (C.S.I.C.),
Apartado 3004 , E-18080 Granada, Spain
\and
    Institut d'Astrophysique de Paris, CNRS, 98bis Bd Arago,
F-75014 Paris, France
\and
    Departamento de Astronom\'\i a, Universidad de Chile, Casilla 36D,
Santiago, Chile
\and
    Observatories of the Carnegie Institution of Washington, 813 Barbara
Street, Pasadena, CA91101
}
\date{Received / Accepted } 
\authorrunning{M\'arquez et al.}
\titlerunning{Long slit spectroscopy of Spirals with and without AGN}

\abstract {We present the kinematical data obtained for a sample of
active (Seyfert) and non active isolated spiral galaxies, based on
long slit spectra along several position angles in the H$\alpha$ line
region and, in some cases, in the Ca triplet region as well. Gas
velocity distributions are presented, together with a simple circular
rotation model that allows to determine the kinematical major
axes. Stellar velocity distributions are also shown. The main result
is that active and control galaxies seem to be equivalent in all
kinematical aspects. For both subsamples, the departure from pure
circular rotation in some galaxies can be explained by the presence of
a bar and/or of a spiral arm. They also present the same kind of
peculiarities, in particular, S-shape structures are quite common near
the nuclear regions.  They define very similar Tully-Fisher relations.
Emission line ratios are given for all the detected HII regions; the
analysis of the [NII]/H$\alpha$ metallicity indicator shows that
active and non-active galaxies have indistinguishable disk
metallicities.  These results argue in favour of active and non-active
isolated spiral galaxies having essentially the same properties, in
agreement with our previous results based on the analysis of near
infrared images. It appears now necessary to confirm these results on
a larger sample.
} \maketitle

\section{Introduction}

The DEGAS collaboration was started a few years ago to search for a
sufficient condition for a galaxy to host an active nucleus. The
active galaxies have been chosen with the following criteria: (a)
catalogued as Seyfert 1 or 2 in V\'eron-Cetty \& V\'eron's catalogue
(we have checked that they were all classified as Seyferts in the
NED database and give their AGN type in Table~\ref{pargal}); (b) with
morphological information in the RC3 catalogue (de Vaucouleurs et
al. 1991); (c) isolated, in the sense of not having a companion within
0.4 Mpc (H$_0$ = 75\kms~Mpc$^{-1}$) and cz$<$ 500 \kms, or companions
catalogued by Nilson without known redshift; (d) nearby, cz $<$ 6000
\kms; and (e) intermediate inclination (30 to 65 \degr). Note
however that the sample of active galaxies thus defined may not be
representative of the whole AGN population, since the selection
criteria of the V\'eron-Cetty \& V\'eron catalogue are not necessarily
complete. The non-active galaxies have been selected among spirals
verifying the same conditions (b), (c), (d) and (e), and with
morphologies (given by the complete de Vaucouleurs coding, not just
the Hubble type) similar to those of the active spirals. A total of 17
active galaxies and 16 non active galaxies (used as control sample),
with similar distribution of morphological types (including the bar
type), were selected. Since all galaxies were chosen to be isolated
under the criteria defined above, side effects that could be produced
by the gravitational interaction with nearby galaxies are avoided.
The details of the program and sample definition are given in
M\'arquez et al. (1999, 2000), where it has been shown that the
non-active galaxies are well suited to be used as a control
sample. Basic properties of the galaxies of our sample are given in
Table~\ref{pargal}.  The observations we carried out included infrared
and optical imaging and long slit spectroscopy to derive the
kinematical properties of the ionized gas and, for some of them, of
the stellar component. Infrared HST images of our selected galaxies
were also retrieved and analyzed when available.

The data already presented have shown some interesting results: the
analysis of the infrared images indicated that there are no
morphological differences between active and non active galaxies in
our sample. We also found that most galaxies have bars and sometimes
nuclear bars within the large bars. Some of these inner bars were
reported for the first time (M\'arquez et al. 1999, 2000). The full
analysis of NGC~6951, the first galaxy of our sample for which all the
required data became available (P\'erez et al. 2000), led us to
discover the decoupling between the stellar and gas kinematics in the
central regions, and to hint the presence of a younger stellar
population there. Velocity drops were detected in the nuclear regions
of five of our sample galaxies. The analysis of the Calcium triplet
(CaT) lines equivalent widths indicated the presence of a different
stellar population spatially coincident with the kinematical
drops. Further analysis of the HST images of a total of 14 galaxies
with reported velocity dispersion drops (including our five), show
evidence for central elongations in most of them, which could be
related with the mechanism invoked in numerical models to reproduce
such drops (see M\'arquez et al. 2003 and references therein).

The present work includes the kinematical data on the gas and stars
obtained for the galaxies of our sample, together with the emission
line ratios measured in the nuclear and extra-nuclear HII regions of
the same galaxies. The metallicities of these HII regions, estimated
from the observed line ratios, are briefly discussed. A general
discussion based on the DEGAS data will be presented in a forthcoming
paper.

The paper is organized as follows. The observations and data reduction
procedures are presented in Sect.~2. Characteristics of the rotation
curves for each galaxy, together with a discussion of the positions of
our sample galaxies in the Tully-Fisher relation are presented in
Sect.~3.  The properties of nuclear and extra-nuclear HII regions are
described in Sect.~4. The summary and conclusions are presented in
Sect.~5.

\begin{table}[t!]
\caption[]{Main properties of the galaxies}
\label{pargal}
\begin{tabular}{llll }
\hline
Galaxy & type  & m$_B$ & AGN \\
       & (RC3) &(mag)  & type \\
       & (1)   & (2)   & (3) \\
\hline
UGC  1395          & .SAT3.. & 14.18       & 1.9 \\        
IC    184          & .SBR1*. & 14.66       & 2 \\        
IC   1816          & .SBR2P? & 13.83       & 1 \\        
UGC  3223          & .SB.1.. & 13.83       & 1 \\        
NGC  2639          & RSAR1*  & 12.59       & 1.9 \\
IC   2510          & .SBT2*. & 14.00$^{*}$ & 2 \\ 
NGC  3660          & .SBR4.. & 12.29       & 2\\                  
NGC  4507          & PSXT3.. & 12.78       & 2 \\        
NGC  4785          & PSBR3*. & 13.21       & 2 \\        
NGC  5347          & PSBT2.. & 13.16       & 2 \\         
NGC  5728          & .SXR1*. & 12.37       & 2 \\         
ESO139-12          & PSAT4P* & 13.59       & 2 \\         
NGC  6814          & .SXT4.. & 11.85       & 1.5 \\         
NGC  6860          & PSBR3.. & 13.68       & 1 \\         
NGC  6890          & .SAT3.. & 13.05       & 2 \\         
NGC  6951          & .SXT4.. & 11.91       & 2 \\       	
\hline                           	     
NGC   151          & .SBR4..  & 12.23       & \\       	     
NGC  1357          & .SAS2..  & 12.46       & \\       
IC    454          & .SB.2..  & 14.50$^{*}$ & \\           
NGC  2712          & .SBR3*.  & 12.38       & \\       
NGC  2811          & .SBT1..  & 12.66       & \\       
NGC  3571          & PSXT1*.  & 12.99       & \\       
NGC  3835          & .S..2*/  & 13.20       & \\       
NGC  4162          & RSAT4..  & 12.55       & \\       
NGC  4290          & .SBT2*.  & 12.66       & \\       
NGC  6012          & RSBR2*.  & 12.69       & \\       
NGC  6155          & .S?....  & 13.20       & \\       
NGC  7328          & .S..2..  & 13.98       & \\       
\hline		     
\end{tabular}

\noindent
(1) Morphological type taken from the RC3 catalogue;\\
(2) B magnitude as given in the RC3 catalogue;\\
(3) AGN Seyfert type taken from the NED.\\
$^{*}$ Taken from NED\\
\end{table}

\section{Observations and Data Reductions}

The spectroscopic data were collected with four different instruments:
the Boller \& Chivens spectrograph attached to the 1.5m ESO telescope
at La Silla (Chile), the Modular Spectrograph at the 2.5m Dupont
telescope at Las Campanas (Chile), the ISIS Spectrograph at the 4.2m
William Herschel telescope in La Palma (Spain) and the TWIN
Spectrograph attached to the 3.5m telescope in Calar Alto (CAHA,
Spain). The setup and main characteristics of the observations are
given in Table~\ref{obs}.

\begin{table*}[t!]
\caption[]{Observations}
\label{obs}
\begin{flushleft}
\begin{scriptsize}
\begin{tabular}{l l c c c c c c c c}
\hline
\noalign{\smallskip}
\multicolumn {1}{l}{Code}
& \multicolumn {1}{c}{Telescope}
& \multicolumn {1}{c}{Date}
& \multicolumn {1}{c}{Instrument}
& \multicolumn {1}{c}{Spectral}
& \multicolumn {1}{c}{Spectral}
& \multicolumn {1}{c}{Range}
& \multicolumn {1}{c}{Slit}
& \multicolumn {1}{c}{Spatial}
& \multicolumn {1}{c}{Average}\\
 & & & & sampling & resolution & & width & sampling & seeing\\
 & & & & (\AA/pix) & {(\AA)} &  (\AA) & (arcsec) & (arcsec/pix) & (arcsec)\\
\noalign{\smallskip}
\hline\noalign{\smallskip}
ESO96 & 1.5m ESO          & Jan.    1996 & Boller\& Chivens & 0.98& 1.86 &5493 - 7505 & 2    &0.78 & 1\\
LC596 & Dupont            & May 1996     & Boller\& Chivens & 1.0 & 1.10 & 6166 - 7190 & 1.25 &0.56 & 1.5\\
LC896 &                   & Aug.   1996  &                  & 1.0 & 1.29 & 6149 - 7181 & 1.2    &0.67 &    \\
LC898 &                   & Aug.   1998  &                  & 0.99& 1.28 & 5540 - 7558 & 1.2  &0.584& \\
WHT96 & WHT               & Aug.   1996  & ISIS             & 0.39& 0.81 & 4757 - 5141 & 1.03 &0.36 & 1.2\\
      &                   &              &                  &   ''&   '' &  6493 - 6880 &    ''&   ''&  ''\\
      &                   &              &                  &   ''&   '' &  8505 - 8881 &    ''&   ''&  ''\\
WHT99 &                   & Dec. 1999  & ISIS               & 0.44& 0.92 & 3654 - 5187 & 1.03 &0.20 & 1.5\\
      &                   &              &                  &   ''&   '' &5354 - 6894 &    ''&   ''&  ''\\
      &                   &              &                  & 0.39& 0.81 &8476 - 8872 &    ''&0.36 &  ''\\
WHT00 &                   & May    2000  & ISIS             & 0.44& 0.92 &3649 - 5185 & 1.00 &0.20 & 1.3\\
      &                   &              &                  &   ''&   '' &5334 - 6874 &    ''&   ''&  ''\\
      &                   &              &                  & 0.39& 0.81 &8456 - 8852 &    ''&0.36 &  ''\\
CAHA00& 3.5m CAHA         & Jan.    2000 & TWIN             & 0.38& 0.80 &4600 - 5380 & 1.20 &0.56 & 1.2\\
      &                   &              &                  &   ''&   '' &6240 - 7000 & 1.20 &0.56 & 1.2\\
\hline
\end{tabular}
\end{scriptsize}
\end{flushleft}
\end{table*}

Standard IRAF\footnote{IRAF is the Image Analysis and Reduction
Facility made available to the astronomical community by the National
Optical Astronomy Observatories, which are operated by the Association
of Universities for Research in Astronomy (AURA), Inc., under contract
with the U.S. National Science Foundation.}  procedures were used for
the reduction of the spectroscopic data, following the standard steps
of bias subtraction, flat field correction, wavelength calibration
with arc lamps observed before and after the target, atmospheric
extinction correction, and flux calibration using spectroscopic
standards observed through an 8~arcsec wide slit for WHT runs. The sky
background level was determined taking median averages over two strips
on either side of the galaxy signal. The parameters of the lines were
measured with the program {\tt splot}. The errors in
Table~4\footnote{available only in electronic form} have been
calculated by quadratic addition of the photon counting errors and of
the errors on the continuum level determination.

We used cross-correlation techniques as described in M\'arquez \&
Moles (1996) to extract the gas kinematical information. The spatial
section in the 2D spectrum with the highest S/N ratio (excluding the
nuclear region for Seyferts\footnote{Notice that in the nuclear
regions of AGNs there are often several components, which we have not
attempted to fit individually.}) was used as a template for the
cross-correlation. The errors refer to the determination of the
velocity shift with respect to the template spectrum. They are shown
as error bars in the velocity distributions. For the stellar velocity
distributions, the spectra in the CaT region were used, with
cross-correlation techniques (as described in P\'erez et al. 2000, and
references therein), using as templates for the cross-correlation the
observed spectra of several stars. The errors refer to the
determination of the velocity shift with respect to the template
spectrum. For clarity, error bars in the stellar velocity
distributions are not plotted, but they typically amount to 50
\kms. The best S/N data have been used to look for the presence
of nuclear dips in the stellar velocity dispersions, and are presented
in M\'arquez et al. (2003).

The sample of active galaxies is more thoroughly observed (more
PAs, higher S/N). Nevertheless, we do not expect this to produce a
severe bias concerning general properties, since only the data along
the major axes are used to characterize the rotation curves and use
them to derive the Tully-Fisher (TF) relation, and to estimate the
metallicity\footnote{As given by [NII]/H$\alpha$, see below.} of disk
HII regions. With respect to the analysis of central structures, the
same kind of S-shaped deviations seems to be present along all major
axes. Besides, for a number of galaxies (most of them active spirals),
enough position angles have been observed and therefore a guess can be
made on what specific feature (disk or bar) may be the origin of such
deviations (see below).

\begin{table*}[t!]
\caption[]{Detailed log of the spectra}
\label{obsspec}

\begin{flushleft}
\begin{scriptsize}
\begin{tabular}{|l|  c r r|| l| c r r |}
\hline
\noalign{\smallskip}
\multicolumn {1}{|l|}{Galaxy}
& \multicolumn{1}{c}{Run}
& \multicolumn{1}{r}{PA}
& \multicolumn{1}{c}{Exp. Time}
& \multicolumn {1}{||l|}{Galaxy}
& \multicolumn{1}{c}{Run}
& \multicolumn{1}{r}{PA}
& \multicolumn{1}{c|}{Exp. Time}\\
\multicolumn {1}{|l|}{(Active)}
& \multicolumn{1}{c}{}
& \multicolumn{1}{r}{(\degr)}
& \multicolumn{1}{c}{(seconds)}
& \multicolumn {1}{||l|}{(Non-active)}
& \multicolumn{1}{c}{}
& \multicolumn{1}{r}{(\degr)}
& \multicolumn{1}{c|}{(seconds)}\\
\noalign{\smallskip}
\hline\noalign{\smallskip}
UGC 1395   & WHT96  &  35 & 1800,1800 &   NGC 151    & LC898  &  75 & 2400\\
           &        & 125 & 0,3600    &              &        & 151 & 2400\\
           & WHT99  &  80 & 3600,1800 &              &        & 165 & 2400\\
           & LC896  & 135 & 3600      &   NGC 1357   & LC898  &  85 & 1860\\
           & LC898  &  45 & 2400      &   IC 454     & CAHA00 & 110 & 3600,1800\\
           & LC898  & 165 & 2400      &              &        & 140 & 3600,1800\\
IC 184     & WHT99  &   7 & 3600,3600 &               & ESO96  &  50 & 3600\\
           &        &  97 & 3600,1800 &               &        & 110 & 3600\\
           & LC896  &   7 & 3600      &               &        & 140 & 3600\\
           &        &  37 & 2400      &               &        & 170 & 3000\\
           &        &  97 & 3600      &    NGC 2811   & ESO96  &  20 & 3000\\
           &        & 157 & 2400      &               &        &  50 & 2400\\
IC 1816    & LC896  &   0 & 2400      &               &        & 110 & 3600\\
           &        &  90 & 2400      &               &        & 170 & 2400\\
UGC 3223   & WHT99  &  80 & 3600,1800 &    NGC 3571   & LC596  &  94 & 3600\\
           & ESO96  &  50 & 3600      &               &        &  64 & 3600\\
           &        &  80 & 3600      &               &        &   4 & 3600\\
           &        & 110 & 2400      &    NGC 6012   & LC896  &  78 & 2400\\
           &        & 170 & 3600      &               &        & 168 & 2400\\
           & CAHA00 &  20 & 3600,3600 &    NGC 7328   & LC896  &  88 & 2400\\
           &        & 170 & 3600,1800 &               &        & 178 & 2400\\
NGC 2639   & WHT99  &   0 & 1800,1800 &               & LC898  &  58 & 3600\\
           &        &  45 & 3600,1800 &               &        & 118 & 2400\\
           &        & 315 & 3600,1800 &               &        &     &     \\
IC 2510    & ESO96  &  58 & 3600      &               &        &     &     \\
           &        & 118 & 3600      &               &        &     &     \\
           &        & 148 & 3600      &               &        &     &     \\
           &        & 178 & 3000      &               &        &     &     \\
NGC 3660   & WHT99  & 115 & 3600,1800 &               &        &     &     \\
           & ESO96  &  25 & 3600      &               &        &     &     \\
           &        &  85 & 2400      &               &        &     &     \\
           &        & 115 & 3600      &               &        &     &     \\
           &        & 145 & 2400      &               &        &     &     \\
           & CAHA00 &  25 & 3600,1800 &               &        &     &     \\
           & WHT00  &  85 & 3600,1800 &               &        &     &     \\
NGC 4507   & ESO96  &  51 & 3600      &               &        &     &     \\
           &        & 141 & 3340      &               &        &     &     \\
NGC 4785   & LC596  &   9 & 1650      &               &        &     &     \\
           &        &  81 & 3600      &               &        &     &     \\
NGC 5347   & WHT00  & 310 & 3600,1800 &               &        &     &     \\
NGC 5728   & LC896  &   0 & 2400      &               &        &     &     \\
           &        &  35 & 2400      &               &        &     &     \\
           &        &  90 & 3600      &               &        &     &     \\
           &        & 150 & 3600      &               &        &     &     \\
ESO 139-12 & LC596  &  35 & 3600      &               &        &     &     \\
           &        &  65 & 3600      &               &        &     &     \\
           &        & 125 & 3600      &               &        &     &     \\
           & LC896  &   0 & 3600      &               &        &     &     \\
           &        &  30 & 3600      &               &        &     &     \\
           &        &  60 & 2400      &               &        &     &     \\
           &        & 120 & 3600      &               &        &     &     \\
           & LC898  &   5 & 2400      &               &        &     &     \\
NGC 6814   & WHT96  &  30 & 1800,1800 &               &        &     &     \\
           &        & 120 & 1800,3600 &               &        &     &     \\
           & WHT00  &  60 & 1800,1800 &               &        &     &     \\
           & LC896  &  29 & 3600      &               &        &     &     \\
           &        & 119 & 2400      &               &        &     &     \\
           & LC898  &  20 & 2400      &               &        &     &     \\
NGC 6860   & LC896  &   4 & 3600      &               &        &     &     \\
           &        &  34 & 3600      &               &        &     &     \\
           &        &  64 & 3600      &               &        &     &     \\
           &        & 124 & 2400      &               &        &     &     \\
           & LC898  &  16 & 2400      &               &        &     &     \\
NGC 6890   & LC896  &   2 & 2400      &               &        &     &     \\
           &        &  62 & 3600      &               &        &     &     \\
           &        & 122 & 2400      &               &        &     &     \\
           &        & 152 & 3600      &               &        &     &     \\
           & LC898  &  13 & 2400      &               &        &     &     \\
\hline
\end{tabular}

\noindent
For the WHT and CAHA runs the two exposure times correspond to
H$\beta$ and H$\alpha$ respectively, the exposure time for the CaT
being the sum of the two. The CaT WHT00 data for NGC~6814 and all the
CaT CAHA00 data did not have sufficient signal to noise ratio to allow
us to derive the stellar velocity distributions.
\end{scriptsize}
\end{flushleft}
\end{table*}

\section{The shape of the rotation curves}

The gas velocity distributions measured along various slit position
angles for all the galaxies of our sample are displayed in
Figs.~\ref{curvas_u1395} to \ref{curvas_n6155}, with the exception of
NGC~2811 for which the [NII] emission lines are not spatially
extended, and were detected in the nuclear region only (H$\alpha$ is
not detected in emission). The kinematical center and systemic
velocities were determined as in M\'arquez et al. (2002): the
kinematical center was defined as the cross section leading to
symmetrical differences between the two horizontal branches of the
rotation curve. That center corresponds, within our resolution, to the
photometric center, i.e., the maximum in the continuum distribution
along the slit, the biggest differences in position always being
smaller than 2 pixels. The redshift corresponding to that kinematical
center was adopted as the redshift of the system.

The stellar velocity distribution, when available, is shown
superimposed on the gas data. Model rotation curves that better
account for the observed data are shown in the figures as full lines;
a normalized arctangent rotation curve function has been used, $v(R) =
v_0 + \frac{2}{\pi}\ v_c\ arctan(R)$, with $R = (r - r_0)/r_t$, where $v_0$
is the velocity of the center of rotation, $r_0$ is the spatial center
of the galaxy, $v_c$ is the asymptotic velocity and $r_t$ is a radius
that corresponds to the transition region between the rising and the
flat parts of the rotation curve (see M\'arquez et al. 2002). Their
corresponding parameters are given in Table~\ref{parspec1}. These do
not correspond to best fits, but only to the models which by eye fit
best the data points along the various PA, with priority given to the
major axis. This model has been used for the sake of comparison
with M\'arquez et al. (2002, hereafter M111) and is quite appropriate
to describe the general shape of the rotation curves of spiral
galaxies, in particular concerning the parameters used for
Tully-Fisher studies (Courteau 1997; M111 and references
therein). These models are also used to constrain the kinematical
major axes, for which the use of a more refined mathematical form is
not critical. In two cases (UGC~3223 and ESO~139-12) the kinematical
major axis is found to be different from the photometric major axis
(taken from the RC3 catalogue).  For UGC~1395, IC~1186 and NGC~4507,
no photometric PA is given in the RC3 (see Table~\ref{parspec1} for
the major axis PA derived from our kinematical data). In a number of
cases the number of slit positions and data points were not sufficient
to obtain a full kinematical model (see below).

\begin{figure}[]
\psfig{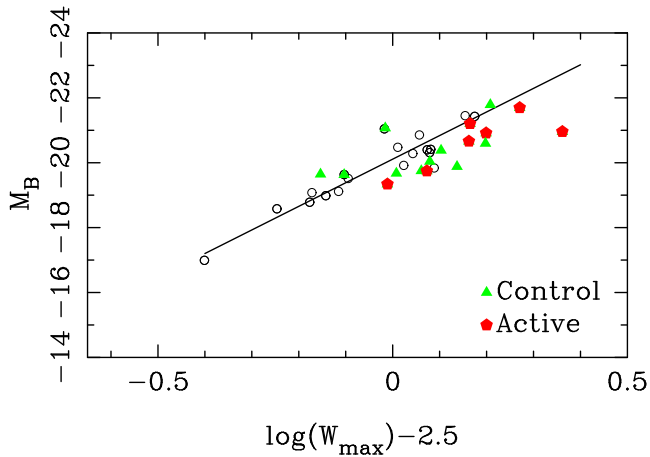}
\caption[]{Tully-Fisher relation for the galaxies of our sample (red
pentagons for active galaxies and green triangles for non-active
ones), together with the isolated galaxies of the M111 sample (empty
circles). M$_B$ is the B corrected absolute magnitude, and W$_{max}$ =
2$\times$V$_{max}$(km/s) is the total amplitude of the rotation
curve. The line indicates the TF relation by Tully \& Pierce (2000).}
\label{tf_tout}
\end{figure}

In order to check whether the galaxies of the DEGAS sample obey the
Tully-Fisher (TF) relation, we plot in Fig.~\ref{tf_tout} the absolute
magnitude B$_{T0}$ (obtained from the corrected B magnitude given in
the RC3) as a function of the rotation amplitude.  The latter was
estimated at R$_{opt}$$\approx$ 1.2$\times$R$_{25}$. The extrapolation
of the model curve was used to get V$_{opt}$= V(R$_{opt}$), as in
M111. It is clear from the figure that active and control galaxies
populate the same area of the TF plane, indicating that their overall
kinematical properties are similar. The small offset along the
x-axis indicates that, within 1$\sigma$, Seyfert galaxies have median
rotation velocities higher that control galaxies, for the same median
luminosity. We also see that they are defining, within 1$\sigma$, the
same TF relation as the isolated galaxies discussed in M111 (the DEGAS
sample comprises earlier spirals, that tend to be more luminous and
therefore are expected to be placed in the upper region of the
diagram). We notice however that, even within 1$\sigma$, most of the
active and control galaxies discussed here are located below the TF
line defined in M111. The reason could be the differences in
morphological content between the samples in M111 and here, or the
need for more accurate absolute magnitudes than those provided by the
RC3. Indeed, it also could be just an artifact of the small number
statistics.

We give below details on the velocity curves of each galaxy with
indications of particular aspects such as departures from the general
rotation curve or differences between the gas and star velocity
distributions. Most galaxies of our sample show rotation curves in
global agreement with circular motions, but including some interesting
details. S-shape structures are quite common near the nuclear regions,
sometimes with different amplitudes on either side, and often showing
a velocity dip along the minor axis. To better see the behaviour
of such departures, the model rotation curve has been subtracted from the data
for each PA and the corresponding residuals have been inspected. 
In fact, zero velocity residuals are expected along the major axis
when an inner disk axis has the same major axis than the large scale
one; a rotation-like sinusoidal feature is expected when these two
axes are misaligned, with amplitudes differing for the different
PAs. This allows hence a better discrimination between nuclear bar or
nuclear disk as the origin of such S-shape features, although for a
definitive conclusion integral field spectroscopy is required.  The
departure from simple rotation explained for some galaxies by the
presence of a bar and/or of a spiral arm are usually also based
on our previously published optical (M\'arquez \& Moles 1996) or
infrared (M\'arquez et al. 1999) images.

\subsection{Active galaxies}

UGC~1395 (Fig.~\ref{curvas_u1395}) has a rather smooth rotation
curve. Only at PA=45\degr~ and 135\degr~ the observed velocity
distribution slightly departs from the overall rotation pattern. There
is no entry for the photometric major axis in the RC3.  Our data
indicate that the kinematical major axis is at PA $\simeq$
165\degr. The gas and stellar rotation curves look pretty similar
along PA = 80\degr~ and 125\degr; but along PA=35\degr~ whereas the
stars follow very well the model $\pm 3$~arcsec from the nucleus, the
gas velocity distribution deviates from this curve by about 30\kms~
2~arcsec NE of the nucleus (see also Fig. \ref{res_cen_u1395}). This
occurs at the position of the bent dust lane structure originating at
the NE extremity of the nuclear bar (Pogge \& Martini 2002). The
residual velocity plots indicate that the S-shape structure interior
to $\pm$5 arcsec, more clearly visible along PA = 125\degr~ and
PA=35\degr, may originate in the inner bar structure (see
Fig. \ref{res_cen_u1395}).

The kinematical and photometric major axes in IC~184 are coincident.
The gas (Fig.~\ref{curvas_i184}) shows a strong asymmetric departure
from the model rotation curve, specially along PA=7\degr~ and 97\degr~
(major and minor axis, respectively) out to $\pm$10~arcsec from the
center, while the stellar velocities follow quite well the model along
PA=97\degr~(the stellar velocity distribution along PA=7\degr~ is
given in M\'arquez et al. 2003).  This is illustrated in the enlarged
figure of the central zone along PA=97\degr~ and in Fig.~1 of
M\'arquez et al. (2003). For the gas, the $\approx$ 100 \kms~
difference between the data points and the rotation model N and NE of
the nucleus along PA=7\degr~ and 37\degr~ are most probably due to the
bar and north spiral arm; the bump observed along PA=97\degr~ is most
probably also due to the bar. Note that a drop in the stellar velocity
dispersion is observed in this galaxy (M\'arquez et al. 2003).  The
residuals of the gas velocity distributions along PA=37\degr~ and
PA=157\degr~ in the inner $\pm$1 arcsec (see Fig. \ref{res_cen_i184})
hint for the presence of a nuclear disk, as those found in some active
galaxies (see M\'arquez et al. 2003 and references therein). Although
IC~184 has been selected as an isolated spiral following the criteria
defined for the DEGAS sample, we recently came upon a paper indicating
that it has several spectroscopically confirmed faint companions
(Kollatschny \& Fricke 1987).

The ionized gas in IC~1816 (Fig.~\ref{curvas_i1816}) presents a rather
smooth velocity distribution along the two perpendicular PAs. Our data
indicate that the kinematical major axis is at PA $\simeq$
150\degr. No value is given in the RC3 for the photometric major
axis. The measured velocities north of the nucleus along PA=0\degr~
are about 50 \kms~ higher than expected, probably due to the presence
of a spiral arm. The small counter-rotation observed roughly $\pm
3$~arcsec from the nucleus could be due to the presence of the inner
bar, but it could also trace the presence of a nuclear disk (see
Fig. \ref{res_cen_i1816}).

UGC~3223 (Fig.~ \ref{curvas_u3223}) presents a rather smooth gas
velocity distribution with a major axis in better agreement with PA
$\simeq$ 70\degr~ than with the photometric major axis given in the
RC3 (PA=80\degr). An S-shaped structure is also visible in this
galaxy, specially along PA=20\degr~ and also along PA=170\degr~ (close
to the kinematical minor axis) where it is asymmetric in
amplitude. The stellar and gas velocity curves in the central region
along PA=80\degr~ are given in Fig.~2 of M\'arquez et al. (2003),
where the presence of a drop in the stellar velocity dispersion is
reported. An S-shape structure at $\pm$ 4 arcsec in the residuals,
observed in all position angles but more clearly visible at PA =
20\degr, can be attributed to the inner bar (see
Fig. \ref{res_cen_u3223}).

NGC~2639 (Fig.~\ref{curvas_n2639}) shows smooth gas and star rotations
along the three PA for which we have data. Along PA=135\degr~ (the
photometric major axis is at PA=140\degr), the stars follow the model
throughout while the gas shows some wiggles (see
Fig. \ref{res_cen_n2639}). Along PA=0\degr, the gas and stellar
velocities follow each other, with an apparent counter-rotation in the
innermost $\pm 1$~arcsec, which is also visible for the gas along
PA=45\degr. The counter-rotation is more clearly evidenced in the
residuals, which show that it is also present along PA=135\degr (see
Fig. \ref{res_cen_n2639}). The rotation curves are given for the gas
and stars along PA=45\degr~ in M\'arquez et al. (2003), where the
presence both of a nuclear disk from HST images and of a drop in the
stellar velocity dispersion are discussed. The residuals in the inner
$\pm$ 2 arcsec seem to be due to the inner bar.

IC~2510 (Fig.~\ref{curvas_i2510}) shows rather smooth gas rotation,
with small deviations about $5-12$~arcsec SE of the nucleus along PA =
148\degr~ and $8-15$~arcsec NW of the nucleus along PA=118\degr, both
probably due to the presence of a bar. A small asymmetric S-shape
structure is present in the central $\pm$3~arcsec, more clearly
detected at position angles other than the major axis in
Fig.~\ref{curvas_i2510}, and along all the PA in the residuals (see
Fig. \ref{res_cen_i2510}). This structure coincides with a region
immediately SW of the major axis, which is an incomplete arc-like
structure visible in the HST image. The global shape of the rotation
curve along the major axis agrees with that given by Persic \& Salucci
(1995).

NGC~3660 (Fig.~\ref{curvas_n3660}), for which the kinematical and
photometric major axes coincide, shows a stellar rotation pattern that
is well fit by the rotation model along PA=115\degr. The velocity
distribution of the gas shows a bump of about 100~\kms~ in the region
between 3 and 5\arcsec~ to the NW of the center, with residuals around
zero for PA=145\degr (see Fig. \ref{res_cen_n3660}); this could
indicate the presence of a nuclear disk perpendicular to this PA. A
possible stellar counter-rotation is seen in the inner $\pm$ 1~arcsec
along PA = 85\degr, but the stellar data for this PA are very noisy
and no firm conclusion can be extracted. The outer parts of the
rotation curve of the gas are not well accounted for by the model,
probably due to the crossing of various spiral arms.

For NGC~4507 (Fig.~\ref{curvas_n4507}) we estimate a kinematical major
axis at PA$\sim$65\degr~ (no value is given in the RC3 for the
photometric axis). The gas velocity distribution along PA=51\degr~ is
in general well represented by the model, but it is somewhat wiggly
along PA=141\degr, where it shows an asymmetric S-shape structure
with a strong dip (about 100 \kms) between 2 and 10~arcsec SE of the
nucleus (see Fig. \ref{res_cen_n4507}).

NGC~4785 (Fig.~\ref{curvas_n4785}) shows a smooth rotation pattern for
the gas along PA=81\degr~ (the major axis), with a velocity drop about
8~arcsec W of the nucleus, probably due to the bar and/or beginning of
the spiral arm; the velocity field has a number of wiggles along
PA=171\degr, with an S-shape structure in the inner $\pm$10~arcsec,
deeper to the South (see Fig. \ref{res_cen_n4785}); no obvious
morphological feature is seen in this region in our infrared image.

NGC~5347 (Fig.~\ref{curvas_n5347}) has too sparse data to be analyzed
in detail.  Notice however that an apparent counter rotation is
noticed in the central $\pm$5~arcsec along both the photometric major
(PA=130\degr) and minor (PA=40\degr) axes (see the inset shown in
Fig. \ref{res_cen_n5347}). This region coincides with the dust lane
seen in the HST map (Pogge \& Martini 2002).

For NGC~5728 (Fig.~\ref{curvas_n5728}), the photometric and
kinematical major axes are in excellent agreement (PA=0\degr). The gas
kinematics show strong departures from the expected, standard disk
rotation model along the four observed PA. The velocity gradient is
very steep in the few arcseconds around the nucleus. Our data are in
agreement with the previous determination by Schommer et al. (1988),
who derived a position angle for the major axis at 2$\pm$5\degr. Our
better spatial resolution allows us to detect a strong velocity
discontinuity a few arcseconds on either side of the nucleus which is
seen along PA=90\degr~ (corresponding to the minor axis) and PA =
150\degr~ in Fig.~\ref{curvas_n5728} and along all the PAs in the
residuals (see Fig. \ref{res_cen_n5728}). These position angles are close
to those used by Emsellem et al. (2001) to derive stellar velocity
distributions within the central $\pm$5~arcsec; the resulting stellar
kinematics are very smooth, and show no velocity discontinuity. This
seems to show evidence for a massive central gaseous disk, decoupled
from the stars, somewhat similar to the one we found in NGC~6951
(P\'erez et al. 2000). However, the presence of two components within
the inner $\pm$10~arcsec detected with a higher spectral resolution by
Prada \& Guti\'errez (1999) could have an influence on the shapes of
the rotation curves.

ESO139-12 (Fig.~\ref{curvas_e139}) appears to be the only unbarred
galaxy in our subsample of active galaxies, even after analysis of the
infrared images (M\'arquez et al. 1999, 2000) and shows smooth
rotation along all observed PA. The differences between the model and
the observed gas kinematics are minimized when the kinematical major
axis is placed at PA=15\degr, instead of 35\degr~ as given in the
RC3 for the photometric axis. An S-shape structure is detected in the
inner few arcseconds, specially visible along PA=120\degr~ and
125\degr. This could correspond to an incomplete spiral-like feature
detected in the innermost 1\farcs8 in the HST image. The
residuals indicate that this feature may be produced by a nuclear disk
with a minor axis close to PA=60-65\degr (see Fig. \ref{res_cen_e139}).

NGC~6814 (Fig.~\ref{curvas_n6814}) is seen too close to face-on for
any kinematical model to be applied. Note however that an asymmetric
S-shaped structure in the gas velocities is observed in the central
arcseconds of all the spectra, with a dominant northern component
reaching maximum amplitude about 4~arcsec from the nucleus. This
feature may be due to the presence of some structure within the bar,
most probably also related to the drop detected in the stellar
velocity dispersion (M\'arquez et al. 2003). Notice that the stars
follow the gas along PA=30\degr~ SW of the nucleus but not in the
other direction. Stellar and gas rotation curves along PA=120\degr~
are also presented in M\'arquez et al. (2003), where the presence of a
broad drop in the stellar velocity dispersion is discussed.

NGC~6860 (Fig.~\ref{curvas_n6860}) has kinematical and photometric
major axes in excellent agreement and shows smooth gas rotation along
its major axis but a number of features, most probably due to the
presence of the bar, are observed along the other PAs, such as an
asymmetric S-like feature close to the nucleus, with departures from
circular rotation up to 100~\kms~ along the minor axis (PA =
124\degr). The velocity gradient in the central region along PA =
64\degr~ (which crosses the bar) is steeper than expected from a
simple rotation model. The residuals along the observed PAs hint
for the structure in these inner $\pm$ 4 arcsec to be due to the inner
bar (see Fig. \ref{res_cen_n6860}).

NGC~6890 (Fig.~\ref{curvas_n6890}) shows quite a regular circular gas
rotation except along its minor axis (PA=62\degr) where a wide
velocity dip is observed NE of the nucleus, corresponding to one of
the sides of an asymmetric S-like feature which is present along
the other PAs (this is more clearly seen in the residuals,
Fig. \ref{res_cen_n6890}). The photometric and kinematical major axes
are the same for this galaxy.

\subsection{Non-active galaxies: new data}

NGC~151 (Fig.~\ref{curvas_n151}) shows overall regular circular
rotation along its major axis at PA=75\degr~ (coincident with the
photometric axis) but there are velocity dips along PA=151\degr~ and
165\degr~ just NW of the nucleus. These dips correspond to an
asymmetric S-shaped structure, with a maximum amplitude of almost
$\pm$ 200 \kms~ along PA=151\degr, {which is seen in the residuals for
all the PAs (see Fig. \ref{res_cen_n151})}. Such kinematical behavior
could be created by a secondary bar 2\farcs5 long at PA=68\degr~
visible in the HST image (see Fig. 34 in M\'arquez et al. 1999).

NGC~1357, an optically unbarred galaxy as classified in the RC3, shows
an overall smooth gas rotation along PA=85\degr, corresponding to the
photometric major axis (Fig.~\ref{curvas_n1357}). A S-like feature is
visible mostly in the left side of the curve (see
Fig. \ref{res_cen_n1357}).

IC~454, for which we determine a kinematical major axis coincident
with the photometric one (PA=140\degr~ as in the RC3), shows somewhat
wiggly gas rotation curves along all the four observed PA
(Fig.~\ref{curvas_i454}). An asymmetric S-like feature is also present
in this galaxy, most clearly visible along PA=110\degr~ in
Fig.~\ref{curvas_i454} but present in the residuals for all PAs (see
Fig. \ref{res_cen_i454}). It produces a dip of about 100 \kms~ in the
velocity distribution, SW of the nucleus along the minor axis
(PA=50\degr). The residuals seem to indicate that this feature is due
to the nuclear bar.

NGC~2811 shows weak H$\alpha$ in absorption and [NII] in emission in
the very central region, and no emission lines further out, so it is
not possible to analyze its kinematics.

NGC~3571 (Fig.~\ref{curvas_n3571}) shows weak extended emission, only
detectable along PA=94\degr~ (its photometrical major axis) with an
overall regular gas rotation. Departures from the model are too noisy
to extract any conclusion (Fig. \ref{res_cen_n3571}).

For NGC~6012 the best model (Fig.~\ref{curvas_n6012}) requires the
kinematical major axis to be at PA=150\degr, as determined from
near-infrared data (M\'arquez et al.  1999), whereas a value of PA =
168\degr~ is given in the RC3. The gas kinematics are compatible with
overall circular rotation along PA=168\degr~ except for a velocity
drop about 30~arcsec S of the nucleus. However, the velocity
distribution is less regular along PA=78\degr, with an S-shape
structure giving rise to a strong and broad dip $0-15$~arcsec W of the
nucleus. The residuals seem to hint for a nuclear disk as the origin
of such motions (see Fig. \ref{res_cen_n6012}). The simple rotation
model used here does not account well for the data along this PA.

NGC~7328 (Fig.~\ref{curvas_n7328}) shows regular circular gas rotation
along the major axis (PA=88\degr, both kinematical and
photometric). Along the minor axis (PA=178\degr) and at PA = 118\degr~
a velocity dip is observed $0-12$~arcsec N of the nucleus,
corresponding to the most conspicuous side of an asymmetric S-like
feature in the central $\pm$10~arcsec. The inspection of the residuals
favour the case of a nuclear disk with a minor axis close to 58\degr~
(see Fig. \ref{res_cen_n7328}).

\subsection{Non-active galaxies: previous data}

The gas kinematics of five galaxies from the DEGAS control sample have
previously been observed along their major axis by M\'arquez \& Moles
(1996). These are: NGC~2712, NGC~3835, NGC~4162, NGC~4290 and
NGC~6155. We made a rotation model similar to that described above and
give the results for these five galaxies in Table~\ref{parspec1}.
Their rotation curves are given in Figs.~\ref{curvas_n2712} to
\ref{curvas_n6155}. For NGC~2712, the rotation curve agrees with that
of H\'eraudeau \& Simien (1998). The residuals with respect to
the model are given in Figs. ~\ref{res_cen_n2712} to
\ref{res_cen_n6155}. In all of them evidences of the presence of
S-shaped central structures can be seen.

\begin{table*}[t!]
\caption[]{Parameters derived from the velocity distributions along the major
axes}
\label{parspec1}
\begin{flushleft}
\begin{scriptsize}
\begin{tabular}{l r r r r r r r r r r r r r }
\hline
\noalign{\smallskip}
\multicolumn {1}{l}{Galaxy}
& \multicolumn {1}{c}{i}
& \multicolumn {1}{c}{D$_{25}$}
& \multicolumn {1}{c}{PA$^{RC3}_{maj}$}
& \multicolumn {1}{c}{PA$_{maj}^{kine}$}
& \multicolumn {1}{c}{r$_G$}
& \multicolumn {1}{c}{v$_G$}
& \multicolumn {1}{c}{r$_1$}
& \multicolumn {1}{c}{v$_1$}
& \multicolumn {1}{c}{r$_m$}
& \multicolumn {1}{c}{v$_m$}
& \multicolumn {1}{c}{cz}
& \multicolumn {1}{c}{r$_0$}
& \multicolumn {1}{c}{v$_0$}\\
~ & & & & & ('') &(km/s) &('') &(km/s) &('') &(km/s) & (km/s) & ('') & (km/s) \\
\noalign{\smallskip}
~ & (1) & (2) & (3) & (4) & (5) & (6) & (7) & (8) & (9) & (10) & (11) & (12) & (13) \\
\noalign{\smallskip}
\hline\noalign{\smallskip}
UGC  1395        & 0.10 & 1.10  & -  & 165  & 1.79 &  29 &  8.8 &  78 &  34 & 140 & 5139 & 4.5 & 132 \\
IC    184        & 0.31 & 1.02  & 7  &   7  & 0.54 & 114 &  9.0 & 127 &  28 & 163 & 5332 & 0.5 & 120 \\
IC   1816        & 0.07 & 1.16  & -  & 150  &      &     &  3.0 & 177 &  21 & 159 & 5095 & 1.5 & 130 \\
UGC  3223        & 0.25 & 1.15  & 80 &  68  & 0.80 &  77 &  1.6 & 111 &  21 & 250 & 4330 & 4.5 & 158 \\
NGC  2639        & 0.22 & 1.26  &140 & 135  &      &     &  6.0 & 229 &  33 & 310 & 3177 & 5.5 & 291 \\
IC   2510        & 0.27 & 1.10  &148 & 148  &      &     & 12.0 & 110 &  27 & 130 & 2529 & 7.5 & 119 \\
NGC  3660        & 0.09 & 1.43  &115 & 115  & 1.50 &  50 & 19.0 & 110 &  64 & 130 & 3441 & 4.5 & 155 \\
NGC  4507        & 0.10 & 1.22  & -  &  65  & 2.10 & 174 &  5.0 & 240 &  36 & 220 & 3689 & 1.5 & 130 \\
NGC  4785        & 0.31 & 1.29  &81  &  81  & 0.93 & 123 &  8.9 & 254 &  50 & 257 & 3672 & 1.5 & 197 \\
NGC  5347        & 0.10 & 1.23  &130 &130?  &    - &   - &    - &   - &   - &   - & 2382 &   - &   - \\
NGC  5728        & 0.24 & 1.49  & 0  &   0  & 0.30 &  85 &  5.0 & 194 &  56 & 189 & 2813 & 1.5 & 180 \\
ESO139-12        & 0.11 & 1.20  &35  &  15  &      &     & 10.0 & 218 & 114 & 252 & 5139 & 5.5 & 180 \\
NGC  6814        & 0.03 & 1.48  & -  &  -   &    - &   -  &   - &   - &   - & -   & 1556 &   - &   - \\
NGC  6860        & 0.25 & 1.13  &34  &  34  & 0.98 &  38 & 10.0 & 184 &  40 & 190 & 4437 & 3.0 & 157 \\
NGC  6890        & 0.10 & 1.19  &152 & 152  & 1.70 &  55 &  6.0 & 120 &  54 & 124 & 2528 & 2.5 & 148 \\
NGC  6951        & 0.08 & 1.59  &170 & 170  &      &     &  7.5 & 145 &  96 & 153 & 1414 & 2.0 & 130 \\
\hline           
NGC   151        & 0.34 & 1.57  &75  &  75  & 0.90 &  78 &  7.5 & 280 & 100 & 227 & 3727 & 1.5 & 170 \\
NGC  1357        & 0.16 & 1.45  &85  &  -   &      &     &  5.5 & 160 &  55 & 137 & 2033 & 2.5 & 115 \\
IC    454        & 0.28 & 1.24  &140 & 140  & 0.40 &  48 & 16.0 & 227 &  44 & 228 & 3973 & 3.5 & 155 \\
NGC  2712$^{**}$ & 0.26 & 1.46   & 178& -  & 2.5 & 106 & 20 & 150 & 49 & 152 & 1833 & 1.5 & 115 \\
NGC  2811        & 0.46 & 1.40  & 20 &   -  &    - &   -  &   - &   - &   - & -   & 2514 &   - &   - \\
NGC  3571        & 0.46 & 1.48  &  94&   -  & 0.90 &  85 &  2.0 & 143 &  26 & 153 & 3771 & 0.8 & 106 \\
NGC  3835$^{**}$ & 0.39 & 1.29 &60&  -  & 0.90 & 32 & 25 & 190 & 48 & 198 & 2452 & 6.5 & 150 \\
NGC  4162$^{**}$ & 0.22 & 1.37 &174& -  & 1.3  & 61 & 22 & 159 & 48 & 160 & 2542 & 2.5 & 130 \\
NGC  4290$^{**}$ & 0.16 & 1.37 &90&  -  & 0.80 & 35 & 11 & 180 & 31 & 180 & 3035 & 3.5 & 145 \\
NGC  6012        & 0.14 & 1.32  & 150 & 168  & -    & -  & 13.6& 99 & 64 & 111 & 1988 & 8.2 & 150\\
NGC  6155$^{**}$ & 0.16 & 1.13 &145& -  & 1.5  & 30 & 12 & 80  & 41 &  90 & 2429 & 5.5 &  95 \\
NGC  7328        & 0.44 & 1.31  &  88&  88  & 2.0  &  73 &  8.0 & 150 &  77 & 150 & 2793 & 2.0 & 107 \\
\hline
\end{tabular}
\null

\noindent
(1) B magnitude as given in the RC3 catalogue;
(2) and (3) inclination and isophotal diameter, respectively, as given in RC3 (in its units);
(4) major axis position angle as given in RC3;
(5) kinematical major axis as determined from our data;
(6) distance to the center of the point out to which the solid boly rotation is seen;
(7) the velocity measured at r$_G$;
(8) distance to the center of the point where the first maximum peak in velocity;
(9) the velocity measured at r$_1$;
(10) maximum distance to the center in the velocity distribution;
(11) the velocity measured at r$_m$;
(12) systemic redshift as measured from our data;
(13) and (14) parameters for the model rotation curve (see text)\\
$^{*}$ Taken from NED\\ 
$^{**}$ Excepting the last two columns, the values are from M\'arquez et al. (1996).
\end{scriptsize}
\end{flushleft}
\end{table*}

\section{Nuclear and disk spectral characteristics}

The spectra for the different HII regions along the slit have been
extracted and the lines measured as explained in M111, producing the
data given in Table~\ref{parhii}\footnote{The columns of this Table
are: 1. galaxy name; 2. slit position angle, anti-clockwise from
north; 3. distance to the nucleus along the slit in arcsec;
4. H$\alpha$ equivalent width; 5 and 6. [OI]$\lambda$6300/H$\alpha$
line intensity ratio and corresponding error; 7 and
8. [NII]$\lambda$6584/H$\alpha$ intensity ratio and corresponding
error; 9 and 10. [SII]$\lambda$6717+6731/H$\alpha$ intensity ratio and
corresponding error; 11. [SII]$\lambda$6717/[SII]$\lambda$6731
intensity ratio.} (available in electronic form only).

Plots of the nuclear spectra for all the galaxies of the DEGAS sample
are displayed in Figs.~\ref{centros1}a-c. The spectra of active nuclei
in general agree with the AGN classification of the V\'eron-Cetty \&
V\'eron (2001) catalogue. Regarding the control galaxies, we find that
5 out of 12 have HII-like nuclei, a fraction that is lower than in
M111. This could be due to the differences in morphological content
between the two samples. We remember here that the control subsample
was defined to match the morphological distribution of the active
galaxies, which are preferentially seen in early type galaxies
(Melnick et al. 1987). For the remaining 7 galaxies in the control
sample, NGC~6012 has been previously classified as a LINER, and the
others present a too strong H$\alpha$ absorption for a reliable
classification.  In any case, given the strength of the [NII] lines
seen in some nuclei, the possibility of some of them hosting dwarf-AGN
nuclei cannot be discarded. Here too, the fact that most of the
galaxies are early type spirals could be relevant since, as a matter
of fact, a higher percentage of low AGN activity is expected for early
type spiral galaxies (Ho et al. 1997).

An estimation of the metallicities of the disk HII regions and
their possible gradients were obtained from the [NII]/H$\alpha$
line ratios only.
Denicol\'o et al. (2002) obtained an empirical calibration of the
oxygen abundances based on this ratio, which they showed to be
powerful when analyzing large survey data to rank their metallicities,
but with quite large uncertainties on individual objects, mainly due
to O/N abundance ratio and ionization degree variations.  Here we only
consider it to study global trends of Z from the spectra of HII
regions in our sample
\footnote{Note that we do not refer to the regions very close to the
AGN in active galaxies.}, along the same lines discussed in M111. The
data don't allow us to conclude on each individual galaxy, but can be
used to look for general trends when the population of the disk HII
regions is considered as a whole. To be able to combine data from
different galaxies, we have normalized the galactocentric distances of
the HII regions to $r_{25}$ (as given in the RC3)\footnote{In the
analysis we only include HII regions with H$\alpha$ equivalent widths
larger than 10 \AA, as in M111. In that way we avoid including regions
with strong Balmer absorption, that could induce inconsistencies in
the estimation of the metallicity.}.  A plot of the [NII]/H$\alpha$
ratios as a function of distance to the galaxy nucleus is displayed in
Fig.~\ref{rhii} for the DEGAS sample, together with the values taken
from the isolated spirals (24 galaxies) of the M111 sample.
It should be noticed that even if the total number of HII regions
involved is large (237), they correspond to 15 active galaxies and
10 non-active ones, with unequal weighting for the number of position
angles per galaxy.
In order to give the same weight to each galaxy, independently of the
number of position angles observed, we consider hereafter only the
HII regions along the major axes, in such a way that they can be
directly compared to the isolated galaxies in M111. Our main result is
that disk metallicities of active and control galaxies are the same,
since the differences are found only at a 1$\sigma$ level: the median
values are [NII]/H$\alpha$ = 0.46 $\pm$ 0.07 for active galaxies and
0.39 $\pm$ 0.07 for non-active ones, to be compared with
[NII]/H$\alpha$ = 0.31 $\pm$ 0.07 for the isolated galaxies in M111
covering the same absolute magnitude range, i.e., M$_B$ between --19
and --22.

We find that there is a deficiency of low [NII]/H$\alpha$ values for
the two DEGAS subsamples. If we now consider two families per
subsample, early (Sa to Sb) and late (Sbc to Scd) spirals, we see that
the deficiency is specially clear for late types.  However, the DEGAS
sample is obviously biased towards early type spirals in the case of
active galaxies (see for instance Moles et al. 1995) and by
construction for the control sample, and therefore no firm conclusions
can be reached.

\begin{table*}[h]
\caption[]{Parameters of the HII regions}
\label{parhii}

\begin{flushleft}
\begin{scriptsize}
\begin{tabular}{l| r r r r r r r r r r}
\hline
\noalign{\smallskip}
\multicolumn {1}{l|}{Galaxy}
& \multicolumn{1}{c}{PA$_{obs}$}
& \multicolumn{1}{c}{R}
& \multicolumn{1}{c}{EW(H$\alpha$)}
& \multicolumn{1}{c}{[OI]/H$\alpha$}
& \multicolumn{1}{c}{err}
& \multicolumn{1}{c}{[NII]/H$\alpha$}
& \multicolumn{1}{c}{err}
& \multicolumn{1}{c}{[SII]/H$\alpha$}
& \multicolumn{1}{c}{err}
& \multicolumn{1}{c}{S1/S2}\\
 & (\degr)& ('') & (\AA) & &([OI]/H$\alpha$) & &([NII]/H$\alpha$) & &([SII]/H$\alpha$) &\\
\noalign{\smallskip}
\hline\noalign{\smallskip}
 es139 &    0. &  -55.6 &    145.300 &  --  &  --  &      0.331 &      0.048 &      0.249 &      0.090 &      1.879\\
 es139 &    0. &  -30.2 &     20.370 &  --  &  --  &      0.389 &      0.096 &      0.259 &      0.183 &      0.858\\
 es139 &    0. &  -12.7 &     17.600 &  --  &  --  &      0.369 &      0.049 &      0.256 &      0.092 &      1.438\\
 es139 &    5. &  -16.2 &     10.690 &  --  &  --  &      0.543 &      0.163 &      0.557 &      0.316 &      1.102\\
 es139 &    5. &  -11.6 &     13.950 &  --  &  --  &      0.353 &      0.056 &      0.214 &      0.104 &      0.806\\
 es139 &    5. &    0.0 &     22.070 &  --  &  --  &      1.194 &      0.012 &      0.811 &      0.020 &      0.904\\
 es139 &    5. &    9.9 &     27.660 &  --  &  --  &      0.351 &      0.048 &      0.218 &      0.090 &      0.970\\
 es139 &    5. &   24.4 &     59.550 &  --  &  --  &      0.378 &      0.039 &      0.241 &      0.072 &      1.089\\
 es139 &   25. &  -22.6 &     15.860 &  --  &  --  &      0.404 &      0.099 &      0.328 &      0.188 &      0.809\\
 es139 &   25. &  -19.1 &     45.410 &  --  &  --  &      0.380 &      0.028 &      0.258 &      0.051 &      1.517\\
 es139 &   25. &  -14.5 &     14.980 &  --  &  --  &      0.398 &      0.052 &      0.290 &      0.098 &      1.297\\
 es139 &   25. &    0.6 &     28.030 &  --  &  --  &      1.109 &      0.008 &      0.660 &      0.014 &      1.084\\
 es139 &   25. &   14.5 &     35.400 &  --  &  --  &      0.408 &      0.035 &      0.223 &      0.063 &      1.226\\
 es139 &   25. &   20.3 &     12.970 &  --  &  --  &      0.510 &      0.118 &      0.256 &      0.221 &      0.846\\
 es139 &   25. &   27.8 &     38.550 &  --  &  --  &      0.391 &      0.062 &      0.217 &      0.114 &      1.173\\
 es139 &   30. &  -34.2 &     10.370 &  --  &  --  &      0.609 &      0.100 &      0.619 &      0.192 &      1.672\\
 es139 &   30. &    0.0 &     14.740 &  --  &  --  &      1.189 &      0.013 &      0.621 &      0.022 &      1.165\\
 es139 &   30. &   12.7 &     15.640 &  --  &  --  &      0.343 &      0.041 &      0.248 &      0.077 &      1.152\\
 es139 &   35. &  -21.3 &     26.260 &  --  &  --  &      0.374 &      0.026 &      0.281 &      0.048 &      1.560\\
 es139 &   35. &    0.0 &     19.930 &  --  &  --  &      1.008 &      0.011 &      0.509 &      0.019 &      0.934\\
 es139 &   35. &   10.6 &     11.890 &  --  &  --  &      0.359 &      0.038 &      0.179 &      0.071 &      1.876\\
 es139 &   35. &   21.8 &     23.070 &  --  &  --  &      0.458 &      0.035 &      0.290 &      0.064 &      1.477\\
 es139 &   60. &  -31.5 &     77.130 &  --  &  --  &      0.425 &      0.053 &      0.256 &      0.095 &      1.559\\
 es139 &   60. &  -27.5 &     26.340 &  --  &  --  &      0.412 &      0.079 &      0.227 &      0.149 &      1.207\\
 es139 &   60. &  -22.8 &     35.640 &  --  &  --  &      0.426 &      0.042 &      0.297 &      0.076 &      1.271\\
 es139 &   60. &  -18.1 &     22.950 &  --  &  --  &      0.435 &      0.060 &      0.329 &      0.114 &      2.204\\
 es139 &   60. &    0.0 &     16.240 &      0.107 &      0.012 &      1.172 &      0.015 &      0.506 &      0.025 &      1.077\\
 es139 &   60. &   11.4 &     14.890 &  --  &  --  &      0.423 &      0.057 &      0.206 &      0.105 &      1.768\\
 es139 &   60. &   17.4 &     13.030 &  --  &  --  &      0.583 &      0.083 &      0.331 &      0.153 &      0.944\\
 es139 &  120. &    0.0 &     30.270 &      0.115 &      0.007 &      1.099 &      0.009 &      0.526 &      0.016 &      0.954\\
 es139 &  120. &   12.7 &     13.450 &  --  &  --  &      0.422 &      0.048 &      0.294 &      0.090 &      1.115\\
 es139 &  120. &   18.1 &     15.240 &  --  &  --  &      0.490 &      0.084 &      0.360 &      0.156 &      2.108\\
 es139 &  120. &   22.8 &     18.530 &  --  &  --  &      0.482 &      0.084 &      0.446 &      0.159 &      1.767\\
 ic184 &    7. &   15.4 &     12.300 &  --  &  --  &      0.588 &      0.079 &      0.315 &      0.149 &      1.513\\
 ic184 &   37. &   12.1 &     17.030 &  --  &  --  &      0.663 &      0.119 &      0.335 &      0.216 &      0.986\\
 ic184 &   97. &   -6.0 &     21.330 &  --  &  --  &      0.419 &      0.033 &      0.211 &      0.061 &      1.025\\
 ic184 &   97. &    0.0 &     12.240 &      0.120 &      0.020 &      1.553 &      0.025 &      0.642 &      0.041 &      1.063\\
 ic184 &  157. &   -9.4 &     11.020 &  --  &  --  &      0.447 &      0.048 &      0.176 &      0.089 &      0.991\\
 ic184 &  157. &    0.0 &     13.730 &      0.121 &      0.017 &      1.548 &      0.022 &      0.441 &      0.035 &      0.790\\
 ic184 &  157. &   14.7 &     13.200 &  --  &  --  &      0.476 &      0.050 &      0.213 &      0.092 &      1.623\\
 i1816 &    0. &  -11.4 &     11.530 &  --  &  --  &      0.629 &      0.084 &      0.435 &      0.155 &      1.059\\
 i1816 &    0. &   -8.0 &     24.780 &  --  &  --  &      0.433 &      0.032 &      0.246 &      0.057 &      1.655\\
 i1816 &    0. &    0.0 &     43.260 &      0.210 &      0.005 &      1.159 &      0.007 &      0.503 &      0.011 &      0.971\\
 i1816 &   90. &  -12.1 &     23.580 &  --  &  --  &      0.445 &      0.039 &      0.360 &      0.073 &      1.193\\
 i1816 &   90. &   -7.4 &     16.410 &  --  &  --  &      0.396 &      0.031 &      0.205 &      0.055 &      1.232\\
 i1816 &   90. &    0.0 &     59.270 &      0.190 &      0.004 &      1.296 &      0.006 &      0.570 &      0.008 &      0.933\\
 i1816 &   90. &   10.1 &     18.790 &  --  &  --  &      0.369 &      0.024 &      0.270 &      0.045 &      1.272\\
 n0151 &   75. &  -53.9 &     92.740 &  --  &  --  &      0.393 &      0.051 &      0.309 &      0.095 &      1.579\\
 n0151 &   75. &  -45.2 &     25.430 &  --  &  --  &      0.570 &      0.116 &      0.725 &      0.229 &      0.963\\
 n0151 &   75. &  -33.6 &     21.270 &  --  &  --  &      0.520 &      0.080 &      0.451 &      0.154 &      1.320\\
 n0151 &  151. &  -34.8 &     39.640 &  --  &  --  &      0.327 &      0.054 &      0.399 &      0.104 &      1.282\\
 n0151 &  151. &  -27.8 &     27.520 &  --  &  --  &      0.512 &      0.130 &      0.457 &      0.248 &      1.425\\
 n0151 &  151. &  -23.8 &     82.140 &  --  &  --  &      0.437 &      0.035 &      0.289 &      0.062 &      1.374\\
 n0151 &  151. &   19.1 &     13.470 &  --  &  --  &      0.369 &      0.049 &      0.274 &      0.092 &      1.211\\
 n0151 &  151. &   27.3 &     21.980 &  --  &  --  &      0.528 &      0.077 &      0.425 &      0.144 &      1.860\\
 n0151 &  151. &   36.0 &     96.900 &  --  &  --  &      0.320 &      0.041 &      0.417 &      0.079 &      1.236\\
 n0151 &  165. &  -27.3 &     49.840 &  --  &  --  &      0.342 &      0.040 &      0.452 &      0.077 &      1.199\\
 n0151 &  165. &  -20.9 &     80.270 &  --  &  --  &      0.397 &      0.030 &      0.337 &      0.056 &      1.263\\
 n0151 &  165. &   27.8 &     23.440 &  --  &  --  &      0.540 &      0.106 &      0.461 &      0.198 &      1.698\\
 n2712 &  178. &  -45.5 &     11.000 &  --  &  --  &      0.349 &      0.104 &      0.582 &      0.208 &      0.876\\
 n2712 &  178. &   -5.2 &     10.540 &  --  &  --  &      0.560 &      0.029 &      0.460 &      0.054 &      1.146\\
 n2712 &  178. &    0.0 &     14.550 &  --  &  --  &      0.530 &      0.015 &      0.430 &      0.028 &      1.139\\
 n2712 &  178. &    5.2 &     28.730 &  --  &  --  &      0.378 &      0.012 &      0.290 &      0.022 &      1.218\\
 n3660 &   25. &    0.0 &     24.190 &  --  &  --  &      0.702 &      0.047 &      0.229 &      0.095 &      1.347\\
 n4785 &   79. &   -9.5 &     10.890 &  --  &  --  &      0.466 &      0.082 &      0.345 &      0.152 &      1.070\\
 n4785 &   79. &    7.3 &     12.440 &  --  &  --  &      0.407 &      0.021 &      0.216 &      0.039 &      1.245\\
 n4785 &   79. &   14.6 &     28.420 &  --  &  --  &      0.441 &      0.022 &      0.331 &      0.041 &      1.438\\
 n4785 &   79. &   19.0 &     21.440 &  --  &  --  &      0.545 &      0.149 &      0.380 &      0.285 &      1.690\\
 n4785 &   89. &   -0.6 &     10.810 &  --  &  --  &      1.855 &      0.030 &      0.708 &      0.046 &      1.094\\
 n4785 &   89. &    0.0 &     12.200 &  --  &  --  &      1.856 &      0.026 &      0.818 &      0.041 &      1.004\\
 n4785 &   89. &    0.6 &     12.620 &  --  &  --  &      1.935 &      0.026 &      0.920 &      0.041 &      0.889\\
 n4785 &   89. &    1.1 &     12.370 &  --  &  --  &      1.984 &      0.029 &      1.078 &      0.046 &      1.597\\
 n4785 &   89. &    1.7 &     11.590 &  --  &  --  &      1.964 &      0.035 &      0.756 &      0.052 &      1.200\\
 n4785 &   89. &    2.2 &     11.880 &  --  &  --  &      1.951 &      0.039 &      0.618 &      0.057 &      1.441\\
 n5728 &   35. &  -69.0 &     17.630 &  --  &  --  &      0.624 &      0.078 &      0.595 &      0.147 &      1.604\\
 n5728 &   35. &  -63.0 &     66.580 &  --  &  --  &      0.479 &      0.023 &      0.274 &      0.041 &      1.341\\
 n5728 &   35. &  -57.6 &     13.740 &  --  &  --  &      0.615 &      0.112 &      0.450 &      0.211 &      1.553\\
 n5728 &   35. &   50.2 &     13.940 &  --  &  --  &      0.555 &      0.121 &      0.263 &      0.226 &      0.912\\
 n5728 &   35. &   52.3 &     16.760 &  --  &  --  &      0.493 &      0.103 &      0.295 &      0.195 &      3.527\\
 n5728 &   35. &   59.0 &     27.000 &  --  &  --  &      0.532 &      0.050 &      0.423 &      0.090 &      1.399\\
 n5728 &   35. &   62.3 &     31.020 &  --  &  --  &      0.439 &      0.042 &      0.329 &      0.078 &      1.391\\
 n5728 &   35. &   67.7 &     16.210 &  --  &  --  &      0.549 &      0.066 &      0.436 &      0.126 &      1.749\\
 n5728 &  150. &   -7.4 &     25.720 &  --  &  --  &      0.830 &      0.024 &      0.615 &      0.043 &      1.504\\
 n5728 &  150. &    0.0 &     50.210 &      0.064 &      0.004 &      1.256 &      0.005 &      0.649 &      0.009 &      1.159\\
 n5728 &  150. &    9.4 &     27.690 &      0.066 &      0.008 &      0.972 &      0.010 &      0.441 &      0.017 &      1.265\\
 
\noalign{\smallskip}
\hline
\end{tabular}
\end{scriptsize}
\end{flushleft}
\end{table*}

\section{Summary and conclusions}

We have presented and discussed the properties derived from the
analysis of long slit spectra along several position angles of a
sample of isolated active and non-active galaxies. A simple model has
been used to trace overall rotation curve properties and obtain the
parameters for Tully-Fisher (TF) analysis. The active and non-active
spirals define very similar TF relations, populating the region
occupied by isolated spirals of early types. In general, the rotation
curves broadly agree with circular rotation, but the pure rotation
model fails to account for large scale velocity distribution
peculiarities, that can be attributed to the presence of a large scale
bar, spiral-arm components, or small-scale central deviations
indicative of the presence of non-circular motions and/or decoupled
disks.  Note that the same kind of small scale central S-shaped
deviations from the rotation curve model are found in both active
(whether they have a single or double barred host, or are of Sy1 or
Sy2 type) and non-active galaxies. When enough position angles are
available, a better indication can be obtained of whether an inner bar
(most cases) or a nuclear disk (see IC~184, NGC~3660, NGC~5728,
ESO~139 and NGC~6012) are at the origin of such features.

Disk HII region metallicities, estimated from the [NII]/H$\alpha$
ratios, are also very similar for the active and non-active
sub-samples.

The analysis of the photometric properties of the DEGAS sample
obtained from NIR imaging has already shown that active and non-active
isolated spiral galaxies share the same disk, bulge and primary bar
properties, and that they are indistinguishable at least down to
spatial scales of about 300 pc (M\'arquez et al. 1999, 2000).  These
results are reinforced here, in the sense that both kinematical
properties and disk metallicities lead to the same conclusion. Thus,
whether or not they have an AGN, the spiral galaxies in the DEGAS
sample are equivalent in all respects (a general discussion based on
the DEGAS data will be presented in a forthcoming paper).  Therefore,
the differences claimed to be related to the mechanisms driving AGN
activity should be found in details related to circumnuclear regions
still unresolved in our observations. But, interestingly, no significant
morphological differences were found, even in the innermost few
arcseconds, for a sample of 123 active and non-active galaxies imaged
with the HST (Martini et al. 2003). Besides, the AGN activity can
involve such different time scales from those of the dynamical
processes under study, that any direct connection would be out of
scope (see e.g. Hunt \& Malkan 1999). Finally, and considering the
growing evidence that all galaxies above a certain luminosity host
central, massive compact objects (McLure \& Dunlop 2002), our results
are fully compatible with the idea that the AGN activity could take
place recurrently in a given galaxy; in this way, the AGN itself would
be seen when fuel is made available in the most internal regions. 
Note however that our sample is small and that differences between the
active and non-active samples may be lost due to the poor statistics
(in the recent works by the SLOAN project, for example, several
thousand galaxies were required to statistically identify clear
differences between active and non active galaxies (Kauffmann et
al. 2003).

\begin{acknowledgements}

I.M. acknowledges financial support from the IAA and the Spanish
Ministerio de Ciencia y Tecnolog\'{\i}a through a Ram\'on y Cajal
fellowship. This work is financed by DGICyT grants PB93-0139,
PB96-0921, PB98-0521, PB98-0684, ESP98-1351, AYA2001-2089,
AYA2001-3939-C03-01 and the Junta de Andaluc\'{\i}a grant TIC-144.  We
acknowledge financial support from the Picasso Programme d'Actions
Integr\'ees of the French and Spanish Ministries of Foreign
Affairs. F.D. acknowledges financial support from CNRS-INSU for
several observing trips. J. Maza gratefully acknowledges support from
the Chilean Centro de Astrof\'\i sica FONDAP 15010003. We thank
Ignacio Marrero for his collaboration in some data acquisition and
calibration. This research has made use of the NASA/IPAC extragalactic
database (NED), which is operated by the Jet Propulsion Laboratory
under contract with the National Aeronautics and Space Administration.
Finally, we are grateful to the referee for a number of interesting
comments that helped us improve the presentation.

\end{acknowledgements}

\begin{figure*}[]
\caption[nucleo1]{Velocity distributions for UGC~1395. As for all
similar figures, gas velocities derived from H$\alpha$ and H$\beta$
are shown as black dots (with error bars) and green circles
respectively. Stellar velocities are shown with blue points.  For
PA=35 and 80\degr~ an enlarged part of the central region is plotted
to see better the corresponding stellar kinematics.  An asterisk after
the PA indicates the slit position angle coinciding with the major
axis. The red line corresponds to the rotation model which best
reproduces the data.}
\label{curvas_u1395}
\end{figure*}

\begin{figure}[]
\caption[nucleo1]{Residuals from the model for UGC~1395. }
\label{res_cen_u1395}
\end{figure}

\begin{figure*}[]
\caption[nucleo1]{Velocity distributions for IC~184. The two sets of
two curves along PA=7 and PA=97 correspond to the LC (left) and WHT
(right) data. An enlarged part of the central zone is shown with 
blue points indicating the stellar component.}
\label{curvas_i184}
\end{figure*}

\begin{figure}[]
\caption[nucleo1]{Residuals from the model for IC~184.}
\label{res_cen_i184}
\end{figure}

\begin{figure*}[htp]
\caption[nucleo1]{Velocity distributions for IC~1816.}
\label{curvas_i1816}
\end{figure*}

\begin{figure*}[htp]
\caption[nucleo1]{Residuals from the model for IC~1816.}
\label{res_cen_i1816}
\end{figure*}

\begin{figure*}[]
\caption[nucleo1]{Velocity distributions for UGC~3223. The two sets of
curves along PA=80\degr~ and PA=170\degr~correspond to 4m class
telescopes (right) and ESO (left) data.}
\label{curvas_u3223}
\end{figure*}

\begin{figure}[]
\caption[nucleo1]{Residuals from the model for UGC~3223.}
\label{res_cen_u3223}
\end{figure}

\begin{figure*}[]
\caption[nucleo1]{Velocity distributions for NGC~2639. The top right
figure shows an enlargement of the central region with the stellar
velocities indicated in blue.  }
\label{curvas_n2639}
\end{figure*}

\begin{figure}[]
\caption[nucleo1]{Residuals from the model for NGC~2639.}
\label{res_cen_n2639}
\end{figure}

\begin{figure*}[]
\caption[nucleo1]{Velocity distributions for IC~2510.}
\label{curvas_i2510}
\end{figure*}

\begin{figure}[]
\caption[nucleo1]{Residuals from the model for IC~2510.}
\label{res_cen_i2510}
\end{figure}

\begin{figure*}[]
\caption[nucleo1]{Velocity distributions for NGC~3660. The ESO96
(left) and WHT99 (right) curves are on the top line for PA=115. A zoom
of the central region of the WHT99 data along PA=115 is shown in the
second line, together with the curve along PA=25 where a few
velocities from the H$\beta$ line are visible in the very center. The
ESO96 (left) and WHT99 (right) curves are on the third line for PA=85.
A zoom of the central region of the WHT00 data along PA=85 is shown in
the fourth line. }
\label{curvas_n3660}
\end{figure*}

\begin{figure}[]
\caption[nucleo1]{Residuals from the model for NGC~3660.}
\label{res_cen_n3660}
\end{figure}

\begin{figure*}[]
\caption[nucleo1]{Velocity distributions for NGC~4507.}
\label{curvas_n4507}
\end{figure*}

\begin{figure}[]
\caption[nucleo1]{Residuals from the model for NGC~4507.}
\label{res_cen_n4507}
\end{figure}

\begin{figure*}[]
\caption[nucleo1]{Velocity distributions for NGC~4785.}
\label{curvas_n4785}
\end{figure*}

\begin{figure}[]
\caption[nucleo1]{Residuals from the model for NGC~4785.}
\label{res_cen_n4785}
\end{figure}

\begin{figure*}[]
\caption[nucleo1]{Velocity distributions for NGC~5347.}
\label{curvas_n5347}
\end{figure*}

\begin{figure}[]
\caption[nucleo1]{Residuals from the model for NGC~5347.}
\label{res_cen_n5347}
\end{figure}

\begin{figure*}[]
\caption[nucleo1]{Velocity distributions for NGC~5728.}
\label{curvas_n5728}
\end{figure*}

\begin{figure}[]
\caption[nucleo1]{Residuals from the model for NGC~5728.}
\label{res_cen_n5728}
\end{figure}

\begin{figure*}[]
\caption[nucleo1]{Velocity distributions for ESO~139-12.}
\label{curvas_e139}
\end{figure*}

\begin{figure}[]
\caption[nucleo1]{Residuals from the model for ESO~139-G12.}
\label{res_cen_e139}
\end{figure}

\begin{figure*}[]
\caption[nucleo1]{Velocity distributions for NGC~6814. An enlargement
of the central $\pm$15 arcsec is given for PA=30\degr~ with the
stellar velocity distribution as blue stars.}
\label{curvas_n6814}
\end{figure*}

\begin{figure*}[]
\caption[nucleo1]{Velocity distributions for NGC~6860.}
\label{curvas_n6860}
\end{figure*}

\begin{figure}[]
\caption[nucleo1]{Residuals from the model for NGC~6860.}
\label{res_cen_n6860}
\end{figure}
\clearpage

\begin{figure*}[]
\caption[nucleo1]{Velocity distributions for NGC~6890.}
\label{curvas_n6890}
\end{figure*}

\begin{figure}[]
\caption[nucleo1]{Residuals from the model for NGC~6890.}
\label{res_cen_n6890}
\end{figure}

\begin{figure*}[]
\caption[nucleo1]{Velocity distributions for NGC~151.}
\label{curvas_n151}
\end{figure*}

\begin{figure}[]
\caption[nucleo1]{Residuals from the model for NGC~151.}
\label{res_cen_n151}
\end{figure}

\begin{figure}[]
\caption[nucleo1]{Velocity distribution for NGC~1357.}
\label{curvas_n1357}
\end{figure}

\begin{figure}[]
\caption[nucleo1]{Residuals from the model for NGC~1357.}
\label{res_cen_n1357}
\end{figure}

\begin{figure*}[]
\caption[nucleo1]{Velocity distributions for IC~454. The two sets of
curves along PA=140\degr~ and PA=110\degr~ correspond to ESO
(left) and CAHA (right) data.}
\label{curvas_i454}
\end{figure*}

\begin{figure}[]
\caption[nucleo1]{Residuals from the model for IC~454.}
\label{res_cen_i454}
\end{figure}

\begin{figure}[]
\caption[nucleo1]{Velocity distribution for NGC~3571.}
\label{curvas_n3571}
\end{figure}

\begin{figure}[]
\caption[nucleo1]{Residuals from the model for NGC~3571.}
\label{res_cen_n3571}
\end{figure}

\begin{figure*}[]
\caption[nucleo1]{Velocity distributions for NGC~6012.}
\label{curvas_n6012}
\end{figure*}

\begin{figure}[]
\caption[nucleo1]{Residuals from the model for NGC~6012.}
\label{res_cen_n6012}
\end{figure}

\begin{figure*}[]
\caption[nucleo1]{Velocity distributions for NGC~7328.}
\label{curvas_n7328}
\end{figure*}

\begin{figure}[]
\caption[nucleo1]{Residuals from the model for NGC~7328.}
\label{res_cen_n7328}
\end{figure}

\begin{figure}[]
\caption[nucleo1]{Average velocity distribution for NGC~2712.}
\label{curvas_n2712}
\end{figure}

\begin{figure}[]
\caption[nucleo1]{Residuals from the model for NGC~2712.}
\label{res_cen_n2712}
\end{figure}

\begin{figure}[]
\caption[nucleo1]{Average velocity distribution for NGC~3835. The
individual points points correspond to the data on either side of the
nucleus.}
\label{curvas_n3835}
\end{figure}

\begin{figure}[]
\caption[nucleo1]{Residuals from the model for NGC~3835.}
\label{res_cen_n3835}
\end{figure}

\begin{figure}[]
\caption[nucleo1]{Velocity distribution for NGC~4162.}
\label{curvas_n4162}
\end{figure}

\begin{figure}[]
\caption[nucleo1]{Residuals from the model for NGC~4162.}
\label{res_cen_n4162}
\end{figure}

\begin{figure}[]
\caption[nucleo1]{Velocity distribution for NGC~4290.}
\label{curvas_n4290}
\end{figure}

\begin{figure}[]
\caption[nucleo1]{Residuals from the model for NGC~4290.}
\label{res_cen_n4290}
\end{figure}

\begin{figure}[]
\caption[nucleo1]{Average velocity distribution for NGC~6155.}
\label{curvas_n6155}
\end{figure}
\clearpage

\begin{figure}[]
\caption[nucleo1]{Residuals from the model for NGC~6155.}
\label{res_cen_n6155}
\end{figure}

\begin{figure*}[]
\caption[nucleo1]{a. Nuclear spectra, plotted as non calibrated flux vs
observed wavelength (in \AA), except for NGC~2639 and NGC~5347 which are 
flux calibrated (in erg s$^{-1}$cm$^{-2}$ \AA$^{-1}$).}
\label{centros1}
\end{figure*}

\begin{figure*}[]
\addtocounter{figure}{-1}
\caption[nucleo1]{b. Nuclear spectra, plotted as non calibrated flux vs
observed wavelength (in \AA).}
\label{centros2}
\end{figure*}

\begin{figure*}[]
\addtocounter{figure}{-1}
\caption[nucleo3]{c. Nuclear spectra, plotted as non calibrated flux vs
observed wavelength (in \AA).}
\label{centros3}
\end{figure*}

\begin{figure}[]
\psfig{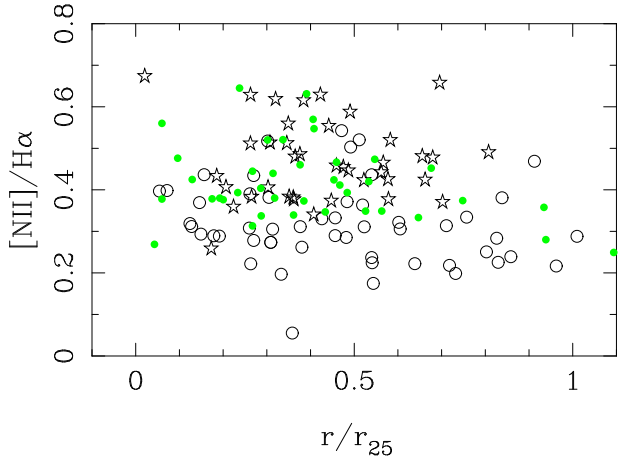}
\caption[]{[NII]/H$\alpha$ intensity ratio as a function of distance
of HII region to the nucleus, normalized to $r_{25}$ for the DEGAS
sample (with stars for active galaxies and green filled circles for
non-active objects) and for the isolated spirals in M111 (empty
circles) having the same absolute magnitude distribution ($-22<M_B <
-19$) (see text).}
\label{rhii}
\end{figure}

\end{document}